\documentclass[12pt,reqno]{amsart}
\usepackage{amsmath,amssymb,amsfonts,amsthm}
\usepackage[mathscr]{eucal}
\usepackage[all]{xy}
\usepackage{hyperref}
\usepackage{setspace}
\usepackage{bm}
\usepackage{xcolor}
\allowdisplaybreaks

\author{D.S.~Kaparulin, S.L.~Lyakhovich, I.A.~Retuntsev}

\address{Physics Faculty, Tomsk State University, Lenin ave. 36, Tomsk 634050, Russia.}

\email{dsc@phys.tsu.ru, \, sll@phys.tsu.ru,\, retuntsev@phys.tsu.ru}

\title[On the world sheet of continuous helicity particle]{On the world sheet of continuous helicity particle}

\begin{document}

\begin{abstract} We consider the class of spinning particle theories, whose quantization corresponds to the
continuous helicity representation of the Poincare group. The
classical trajectories of the particle are shown to lie on the
parabolic cylinder with a lightlike axis irrespectively to any
specifics of the model. The space-time position of the cylinder is
determined by the values of momentum and total angular momentum. The
value of helicity determines the focal distance of parabolic
cylinder. Assuming that all the world lines lying on one and the
same cylinder are connected by gauge transformations, we derive the
geometrical equations of motion for the particle. The timelike world
paths are shown to be solutions to a single relation involving the
invariants of trajectory up to fourth order in derivatives.
Geometrical equation of motion is non-Lagragian, but it admits
equivalent variational principle in the extended set of dynamical
variables. The lightlike paths are also admissible on the cylinder,
but they do not represent the classical trajectories of this
spinning particle. The classical trajectories of massless particle
(with zero helicity) are shown to lie on hyperplanes, whose
spacetime position depends on momentum and total angular momentum.
\end{abstract}

\maketitle

\section{Introduction}
The spinning particles are studied since the work of Frenkel
\cite{Fren}. For review of results before 1968, we cite the book
\cite{Corben}, later studies can be found in \cite{Fryd}. For the
more recent research, we refer to \cite{LSS, Silenko} and references
therein. The concept of spinning particle admits generalization in
low-dimensional space-times \cite{GKL1}, \cite{AL}, and higher
dimensions \cite{LSS6d, LSShd1, LSShd1/2}. For the constant
curvature spaces, we cite \cite{KuLSS1}. It is believed that the
spinning particle models provide the realistic quasi-classical
description of the motion of real elementary particles with spin and
localized twisted wave packets \cite{KI}. For the applications
spinning particle concept in high energy and accelerator physics,
astrophysics we mention \cite{App1, App2, App3, App4, App5, App6,
App7, App8, App9, App10}.

The class of spinning particle models whose quantization corresponds
to the irreducible representation of the Poincare group is of
interest. The examples are given above. The method of
Kirillov-Konstant-Soureau \cite{Kirillov, Kostant, Sour} tells us
that the quantization of the model leads to irreducible
representation if its classical limit is a dynamical system on the
co-orbit of corresponding group. In this setting, the state of the
irreducible particle is determined by the values of momentum $p$ and
total angular momentum $J$, being subjected to mass-shell and
spin-shell conditions. All the gauge invariant observables are the
functions of $p$, $J$. The action functional is given by the
symplectic form on the co-orbit.

The majority of irreducible spinning particle models have one common
feature: the generalized coordinates include, besides the particle
position, the position in internal space. The configuration space of
the model is given by the fiber bundle
$\mathbb{R}^{1,d-1}\times\mathcal{S}$, with the internal space
$\mathcal{S}$ being typical fiber. The Lagrangian of the model is a
function of Lorentz invariant combinations of the generalized
coordinates and their derivatives, typically up to the first order.
The structure of Lagrangian is selected from the requirement of
irreducibility of Poincare group representation. The equations of
motion follow from the least action principle, and they inevitably
involve internal coordinates. In all these models, the relationship
between the representation and classical dynamics is hidden in the
structure of Lagrangian. The universal dynamical principle selecting
the trajectories of spinning particle has been unknown for a long
time.

In the current article, we consider the spinning particle motion
using the recently proposed world sheet concept \cite{LK}. In this
paper, it has been shown that the classical trajectories of the
irreducible spinning particle inevitably lie on the cylindrical
(hyper)surface in Minkowski space.  The (hyper)surface was termed
the world sheet of spinning particle. The shape of the world sheet
is determined by the representation. The world sheet position in the
space-time is determined by the values of the momentum and total
angular momentum, being subjected to the mass shell and spin shell
conditions. The dynamical equations for the particle classical
trajectories follow from the fact that all the world lines that lie
on one and the same world sheet are connected by the gauge
transformations. The resulting equations of motion are purely
geometrical relations on the particle world line, and they do not
involve extra variables.

The geometry of world sheets of massive spinning particles has been
described in the original paper \cite{LK}. The world surfaces has
been shown to be toroidal cylinders of dimension $[(d+1)/2]$ with
timelike axis (the square brackets determine the integer part of
enclosed number). In $d=3,4$, the world surfaces are circular
cylinders with timelike axis. The value of momentum determines the
direction of symmetry axis, and the value of total angular momentum
defines the position of cylinder in Minkowski space. The spinning
particle trajectories are general curves on circular cylinders. The
geometrical equations of motion in $d=3$ have been derived in
\cite{LK}. In $d=4$, the same problem has been solved in
\cite{Ret1}. An alternative derivation of equations of motion for
cylindrical curves has been given in \cite{Ret2}. In all the cases,
the differential equations involve invariants of trajectory up to
the fourth order in derivatives in a very complicated combination,
and they are known only in implicit form. The geometrical equations
of motion are non-Lagrangian, but they admit equivalent variational
formulation with extra fields. This formulation represents
previously known model \cite{GKL1}. The world sheet concept admits
inclusion of interactions with external electromagnetic field
\cite{KR1}. At the interaction level, the spinning particle paths
still lie on a two-dimensional hypersurface, but its shape depends
on the configuration of external field.

In the articles \cite{NerRam1, NerRam2, NerMan}, it has been noted
that the helices with a lightlike tangent vector and timelike
symmetry axis are admissible trajectories for massive particle.  In
the framework of the world sheet formalism, these curves represent
the cylindrical paths with a lightlike tangent vector. The
geometrical equations of motion for lightlike helical paths in
$d=3,4$ derived in \cite{LK}, \cite{Ret1}. It has been shown that
the lightlike helices represent the special class of cylindrical
world lines with a reduced gauge symmetry. In so doing, each helix
forms the gauge equivalence class of its own. This suggests that the
helices can be considered as the one-dimensional spinning particle
world sheets. The construction \cite{NerRam1, NerRam2, NerMan} seems
to be unique in space-time dimension $d=3,4$ and it applies only for
massive particles. We do not know geometrical models of massive
particles with lightlike trajectories in higher dimensions. As for
massless particles, their positions are known to lie on the
hyperplanes \cite{Duval1, Duval2}, but the general fact is not
proven up to date.

In the present article, we apply the world sheet concept for the
study of dynamics of massless spinning particles with continuous
helicity in $d=3$ Minkowski space. The world sheets are parabolic
cylinders with lightlike symmetry axis. The focal distance of the
parabolic cylinder is determined by helicity. The world sheet
position is determined by the values of momentum and total angular
momentum. The geometrical meaning of $p$ and $J$ is explicitly
identified. Assuming that all the particle trajectories lying on one
and the same cylinder are connected by the gauge transformations, we
derive the equations of motion for the curves with timelike tangent
vector on the world sheet. We show that the particle paths enjoy a
single differential equation involving invariants of trajectory up
to the forth order. The equation of motion is non-Lagrangian, but it
admits equivalent variational formulation with extra dynamical
variables, been previously known \cite{GKL1}. The lightlike
trajectories are identified with the paths with zero curvature.
These paths do not correspond to the trajectories of irreducible
particle. The world sheet of the particle with zero mass and
helicity are shown to be hyperplanes. Hence, the trajectories of
such particles are proven to be planar curves.

The article is organized as follows. In Section 2, we describe the
geometry of the world sheet of massless particle with continuous
helicity in three-dimensional Minkowski space. In Section 3, we
derive the equations with hight derivatives for general cylindrical
path on the world sheet. Timelike and lightlike trajectories are
considered. In Section 4, we construct Hamiltonian formulation for
the model and discuss the correspondence with the previously known
model of such a particle. In Section 5, we discuss the dynamics of
massless particles. The conclusion summarizes the results.

\section{Irreducibility conditions and world sheet}
We consider the spinning particle that travels in $3d$ Minkowski
space. The particle position is denoted by $x^{\,\mu}\,,\mu=0,1,2$,
the momentum is $p$ and total angular momentum is $J$. We assume
that the quantization of the model corresponds to the irreducible
representation of the Poincare group with a continuous helicity.
Irreducibility means that the momentum $p$ and total angular
momentum $J$ meet the mass-shell and spin-shell conditions,
\begin{equation}\label{ms-shell}
    \phantom{\frac12\bigg(}(p,p)=0\,,\qquad (p,J)=\sigma\,.\phantom{\frac12\bigg(}
\end{equation}
Here, the round brackets denote the scalar product with respect to
the Minkowski metric. We use a mostly positive signature of the
Minkowski metric \mbox{$\eta_{\,\mu\nu}=\text{diag}(-1,1,1)$}
throughout the paper. Constant parameter $\sigma$ is helicity. The
case $\sigma=0$ corresponds to the massless particle. The values of
momentum $p$ and total angular momentum $J$, being subjected to
condition (\ref{ms-shell}), determine the state of spinning
particle. The space of all the classical spinning particle states is
associated with the co-orbit (\ref{ms-shell}) of the Poincare group.

The vector of spin angular momentum $M$ is determined by the rule
\begin{equation}\label{Mdef}
    \phantom{\frac12\bigg(}M=J-[x,p]\,,\phantom{\frac12\bigg(}
\end{equation}
where the square brackets denote the cross product in $3d$
space-time. We use the convention
\begin{equation}\label{vectprod}
    \phantom{\frac12\bigg(}[u,v]=\epsilon_{\mu\nu\rho}u{}^\mu v{}^\nu dx{}^\rho\,,\qquad\epsilon_{012}=1\,,\phantom{\frac12\bigg(}
\end{equation}
where $\epsilon_{\mu\nu\rho}$ is $3d$ levi-Civita symbol, and $u$,
$v$ are test vectors. In accordance with our definition, the
representation for double cross product of three test vectors $u$,
$v$, $w$ reads
\begin{equation}\label{}
    \phantom{\frac12\bigg(}[u,[v,w]]=w(u,v)-v(u,w)\,.\phantom{\frac12\bigg(}
\end{equation}
We note that the last formula is sensitive to the particular choice
of the signature of the metric. The article \cite{LK} tells us that
the spin angular momentum vector $M$ must be normalized in every
irreducible spinning particle theory,
\begin{equation}\label{L-shell}
    (M,M)=\varrho\,.
\end{equation}
The value of $\varrho$ distinguishes representations with one and
the same value of helicity $\sigma$ (\ref{ms-shell}).

Conditions (\ref{ms-shell}), (\ref{L-shell}) have consequences.
Combining (\ref{L-shell}) with the definition of the spin angular
momentum (\ref{Mdef}), we see that the set of particle positions in
the state with prescribed values of momentum $p$ and total angular
momentum $J$ forms a hypersurface in Minkowski space,
\begin{equation}\label{cylinder}
    \phantom{\frac12}(p,x)^2+2(v,x)+a=0\,.\phantom{\frac12}
\end{equation}
Here, $p$ is the particle momentum, and quantities $v$, $a$ are
determined by the total angular momentum $J$ by the following rule:
\begin{equation}\label{v-a}
    \phantom{\frac12}v=[J,p]\,,\qquad a=(J,J)-\varrho\,.\phantom{\frac12}
\end{equation}
By definition (\ref{v-a}), the momentum $p$ is lightlike, and the
vector $v$ is normalized and orthogonal to $p$,
\begin{equation}\label{v-p-cond}
    \phantom{\frac12}(v,p)=(p,p)=0\,,\qquad (v,v)-\sigma^2=0\,.\phantom{\frac12}
\end{equation}
The hypersurface, being defined by equation (\ref{cylinder}), is
termed a world sheet of spinning particle. By construction, it
includes all the particle positions with the momentum $p$ and total
angular momentum $J$.

The hypersurface, being defined by equation (\ref{cylinder}), is a
parabolic cylinder with lightlike axis. The lightlike vector $p$
determines the direction of symmetry axis. The spacelike vector $v$
determines the direction of asymptotes of parabolas, being
orthogonal sections of cylinder. The focal distance of parabolic
cylinder is determined the helicity. It equals $\sigma$. The
quantity $a$ determines the distance between the vertex of parabolic
cylinder and origin. The cylinder parameters determine the total
angular momentum $J$ by the rule
\begin{equation}\label{J-a}
    J=\frac{a+\varrho}{2\sigma}v+\sigma e\,.
\end{equation}
The relation involves auxiliary lightlike vector $e$ which is
orthogonal to $v$ and has a normalized scalar product with $p$,
\begin{equation}\label{e-p-v}
    \phantom{\frac12}(e,e)=(e,v)=0\,,\qquad (e,p)=1\,.\phantom{\frac12}
\end{equation}

Equation (\ref{cylinder}) has consequences. It shows that the
classical positions of the particle with continuous helicity lie on
a parabolic hypercylinder with lightlike axis in Minkowski space.
The position of hypersurface is determined by the momentum and total
angular momentum. Formula (\ref{v-a}) expresses the values of
cylinder parameters $p$, $v$, $a$ in terms of $p$ and $J$. The
relationship between the quantities $p$, $v$, $a$ and $p$, $J$ is a
bijection. Formula (\ref{J-a}) determines the momentum and total
angular momentum of the particle in terms of cylinder parameters.
This result means that the co-orbit of the particle with continuous
helicity can be parameterized by the set of parabolic cylinders with
lightlike axis, being the world sheets. The relationship between the
world sheets and co-orbit points is purely geometrical. The world
sheet is the hypersurface including the all the possible particle
positions with given values of $p$ and $J$. The inverse is also true
because the space-time position of the world sheet determines the
state of the particle.

World sheet concept determines a dynamical principle that governs
the particle motion. Equation (\ref{cylinder}) represents a single
restriction imposed onto the particle positions that follows from
the irreducibility condition of the Poincare group representation.
This means that all the points of the world sheet represent the
possible particle positions, while the world paths must be general
curves on the world sheet. Assuming that all the curves that lie on
one and the same world sheet are connected by the gauge
transformation, we associate the classical paths of continuous
helicity particle with general cylindrical lines on parabolic
cylinders with a timelike axis. This reduces the task of description
of particle path by differential equations to the problem of
classification of cylindrical curves. We elaborate  on this problem
in the next section.

\section{World paths on world sheets}

\subsection{Problem setting}
In the present section, we consider the problem of classification of
curves on parabolic cylinders with lightlike axis in Minkowski
space. We address three questions: (i) to derive a (system of)
differential equations describing the curves on the parabolic
cylinder; (ii) to express the parameters of the world sheet (and,
hence, the particle state) in terms of derivatives of world path;
(iii) to identify the gauge symmetries of the model. The problem
represents the particular case of more general task of description
of the class of curves lying on the set of surfaces. The solution to
questions (i), (ii) is well-known in the differential geometry of
the curves. We cite the textbook \cite{DiffGeom} for details. The
solution to question (iii) is given in \cite{LK} for the first time,
even though the problem is quite simple in itself.

The description of curves on circular hypercylinder in $3d$
Euclidean space has been first studied in \cite{1966}. The problem
has been recently reconsidered in \cite{C1}, \cite{C2}. It has been
shown that the cylindrical curves are described by a single scalar
equation of fourth order. In $3d$ Minkowski space, the cylindrical
curves has been studied in \cite{LK}. In $4d$ Minkowski space, the
curves on $2d$ circular cylinders are classified in \cite{Ret1}. The
mentioned above articles use different approaches. In the work
\cite{C1}, the concept of constant separation curves is used. The
article \cite{C2} assumes that the cylinder is determined by
algebraic equation. The last approach is best suited for
classification of spinning particle trajectories because the world
sheet is determined by the algebraic equation (\ref{cylinder}).

In our classification of paths on parabolic cylinders we mostly
follow article \cite{C1}, some additional comments are given in
\cite{LK}. As the complete solution to the problem involves many
technical details, we first explain the general method. The
computation details are given in the next subsections. The world
sheet of continuous helicity spinning particle  is determined by the
equation (\ref{cylinder}). The curve $x(\tau)$ lies on the cylinder
(\ref{cylinder}) if the equation of hypersurface is satisfied for
all the values of the parameter $\tau$. This implies infinite set of
differential consequences,
\begin{equation}\label{dc}
    \frac{d^k}{d\tau^k}\Big((p,x)^2+2(v,x)+a\Big|_{x=x(\tau)}\Big)=0, \qquad k=0,1,2,\ldots.
\end{equation}
These relations represent an overcomplete system of equations that
connects the derivatives of trajectory and cylinder parameters. The
relevant information is included into the differential consequences
of orders $k=0,\ldots,4$. It includes five equations for four
independent components of $p$, $v$, $a$ subjected to
(\ref{v-p-cond}). Solving these equations with respect to $p$, $v$,
$a$, we express the parameters of cylinder in terms of parameters of
trajectory. This solves the problem (ii). The consistency condition
for the system (\ref{dc}) is a differential equation, being
satisfied by the cylindrical paths. This equation represents a
solution to the problem (i). By construction, it involves the
derivatives of world path up to fourth order. The higher-order
differential consequences (\ref{dc}) must follow from the lower
order ones, so they are not independent.The gauge symmetries of the
model are generated by the shifts along the cylindrical surface. So,
the gauge generators are associated with the basis vectors in the
tangent space to the cylinder. This solves the problem (iii).

The described above procedure has several subtleties. First, the
world lines representing the classical trajectories of spinning
particles must be causal. The causality condition $\dot{x}^0>0$ is
imposed to prevent existence of closed world loops, which are
considered unphysical. Throughout the article only causal curves are
considered. Second, the tangent vector to the Minkowski space curve
can be timelike, lightlike or spacelike. Our analysis shows that the
cylindrical curves of different type are not connected by the gauge
transformations. So, the classification problems for spacelike,
timelike, and lightlike cylindrical curves represent different
tasks. In the present work, we consider the timelike and lightlike
paths because they automatically meet causality condition. The
spacelike curves can be included in general scheme in a similar way,
but the causality may be an issue.

Finally, the cylinders can intersect. The intersection line belongs
to several cylinders in the set (\ref{cylinder}), so equations
(\ref{dc}) do not have a unique solution with respect to the
parameters $p$, $v$, $a$. These paths must be excluded because they
do not determine a particle state in an unambiguous way.  In the
case of parabolic cylinders set, the intersection is a line with
lightlike tangent vector or non-casual curve. The line appears if
the cylinders with one and the same direction of axis are intersect.
One line belongs to the infinite number of world sheets with one and
the same direction of symmetry axis. The non-casual curve appears if
two cylinders with different directions of symmetry axis intersect.
All the mentioned paths are excluded.

In the paper \cite{LK}, the curves that lie one a unique
representative in the class of world sheets was termed typical. The
curves that belong to multiple world sheets were termed atypical.
The classification of timelike and lightlike lines on parabolic
cylinders presented in subsections 3.1 and 3.2 considers only
typical curves. The atypical curves (including lightlike straight
lines) are systematically ignored below.

\subsection{Timelike world lines on the parabolic cylinder}
Now we can proceed with explicit derivation of the equations of
cylindrical path. The differential consequences of (\ref{cylinder})
up to fourth order have the form
\begin{equation}\label{3}\begin{array}{c}\displaystyle
(\dot{x},n)=0\,,\quad(\ddot{x},n)+(\dot{x},p)^2=0\,,\quad(\dddot{x},n)+3(\dot{x},p)(\ddot{x},p)=0\,,\\[3mm]
(\ddddot{x},n)+4(\dot{x},p)(\dddot{x},p)+3(\ddot{x},p)^2=0\,.
\end{array}\end{equation}
Here, we use a notation,
\begin{equation}\label{pnapva}\begin{array}{c}\displaystyle
n=(x,p)p+v\,.
\end{array}\end{equation}
The new vector $n$ subjects to the conditions
\begin{equation}\label{p-n-cond}\begin{array}{c}\displaystyle
(p,n)=0\,,\qquad (n,n)=\sigma^2\,.
\end{array}\end{equation}
As we can see from the first equation, vector $n$ defines the normal
to tangent space to the world sheet at the point with coordinate
$x$. As far as $n$ is spacelike, the tangent space has the Lorentz
signature in each point of the cylinder.

Let us turn to the description of cylindrical curves. Assume that
$x(\tau)$ is a timelike world line parameterized by the natural
parameter. The velocity vector is normalized,
\begin{equation}\label{timecond}
    \phantom{\frac12}(\dot{x},\dot{x})=-1\,.\phantom{\frac12}
\end{equation}
Throughout the section, the dot denotes the derivative by the
natural parameter $\tau$. The Frenet-Serret moving frame, being
associated with the timelike curve $x(\tau)$, reads
\begin{equation}\label{Frenet-frame}\begin{array}{c}\displaystyle
 e_{0}=\dot{x},\qquad e_{1}=\frac{\ddot{x}}{\sqrt{(\ddot{x},\ddot{x})}}\,,\qquad
 e_{2}=\frac{[\dot{x},\ddot{x}]}{\sqrt{(\ddot{x},\ddot{x})}}\,.
\end{array}\end{equation}
The basis vectors $e_a,a=0,1,2$ of the Frenet-Serret frame are
normalized and orthogonal to each other,
\begin{equation}\label{6}\begin{array}{c}\displaystyle
\phantom{\frac12}-(e_{0},e_{0})=(e_{1},e_{1})=(e_{2},e_{2})=1,\quad
(e_{0},e_{1})=(e_{0},e_{2})=(e_{1},e_{2})=0\,.\phantom{\frac12}
\end{array}\end{equation}
The vector $e_0$ is timelike, and the vectors $e_a,a=1,2$ are
spacelike. Condition (\ref{timecond}) and basis (\ref{Frenet-frame})
are well-defined for each timelike curve, which is not a straight
line. This does not restrict generality because no rectilinear paths
with timelike tangent vector lie on the parabolic cylinder with the
lightlike axis.

The Frenet-Serret formulas for the timelike curve $x(\tau)$ read
\begin{equation}\label{Frenet-formulas}
    \dot{e}_0=\varkappa_1e_1\,,\qquad
    \dot{e}_1=\varkappa_1e_0+\varkappa_2e_2\,,\qquad
    \dot{e}_2=-\varkappa_2e_1\,.
\end{equation}
The curvature $\varkappa_1$ and torsion $\varkappa_2$ of the curve
are determined by the rule
\begin{equation}\label{7}\begin{array}{c}\displaystyle
\varkappa_{1}=\sqrt{(\ddot{x},\ddot{x})}\,,\qquad
\varkappa_{2}=\frac{(\dot{x},\ddot{x},\dddot{x})}{\sqrt{(\ddot{x},\ddot{x})}}
\end{array}\end{equation}
By construction, the curvature $\varkappa_1$ is a positive number,
and the torsion $\varkappa_2$ is a real quantity. With account for
conditions (\ref{Frenet-formulas}), the time derivatives of particle
position can be expressed as the linear combinations of the
Frenet-Serret basis vectors (\ref{Frenet-frame}) with the
coefficients depending on the curvature and torsion of path and
their derivatives,
\begin{equation}\label{8}\begin{array}{c}\displaystyle
\dot{x}=e_0\,,\qquad \ddot{x}=\varkappa_1e_1\,,\qquad
\dddot{x}=\varkappa_1{}^2e_0+\dot{\varkappa}_1e_1+\varkappa_1\varkappa_2e_2\,,\\[5mm]\displaystyle
\ddddot{x}=3\dot{\varkappa}_1\varkappa_1e_0+(\ddot\varkappa_1+\varkappa_1{}^3-\varkappa_1\varkappa{}_2{}^2)e_1+
(2\dot{\varkappa}_1\varkappa_2+\varkappa_2\dot\varkappa_2)e_2\,.
\end{array}\end{equation}
The representation for $\dddot{x}$ involves the derivative of
curvature $\dot{\varkappa}_1$. The representation for $\ddddot{x}$
involves the second derivative of curvature $\ddot{\varkappa}_1$,
and first derivative of torsion $\dot{\varkappa}_2$.

The unknown vectors $p$, $n$ are determined by the conditions
(\ref{3}) and (\ref{p-n-cond}). We seek the solution to
(\ref{p-n-cond}) in the following form:
\begin{equation}\label{10}\begin{array}{c}\displaystyle
p=\gamma\sqrt{\text{sign}(\sigma)\sigma\varkappa_1}(e_0-\beta
e_1+\alpha e_2)\,,\qquad n=\text{sign}(\sigma)\sigma(\alpha
e_1+\beta e_2)\,,
\end{array}\end{equation}
where $\alpha,\beta,\gamma$ are new dimensionless unknowns. The
quantities $\alpha,\beta$ are subjected to the condition
\begin{equation}\label{}
    \alpha^2+\beta^2=1\,.
\end{equation}
The quantity $\gamma$ is positive because $p^0>0$. On substituting
representation (\ref{10}) into (\ref{3}), we arrive at the following
system of algebraic equations for $\alpha$, $\beta$ and $\gamma$:
\begin{equation}\label{11}\begin{array}{c}\displaystyle
\alpha^2+\beta^2-1=0\,,\qquad \alpha+\gamma^2=0\,,\qquad
B\alpha+C\beta+3\gamma^2\beta=0\,,\\[5mm]\displaystyle
E\alpha+D\beta+4\gamma^2(B\beta-C\alpha)+\gamma^2(7-3\alpha^2)=0\,.
\end{array}\end{equation}
Here, the notation is used,
\begin{equation}\label{12}\begin{array}{c}\displaystyle
B=\varkappa_1{}^{-2}\dot{\varkappa_1}\,,\qquad
C=\varkappa_1{}^{-1}\varkappa_2\,,\\[5mm]\displaystyle
D=\varkappa_1{}^{-3}(2\dot{\varkappa_1}\varkappa_2+\varkappa_1\dot{\varkappa_2})\,,\qquad
E=\varkappa_1{}^{-3}(\ddot{\varkappa_1}+\varkappa_1^3-\varkappa_1\varkappa_2^2)\,.
\end{array}\end{equation}
Conditions (\ref{11}), (\ref{12}) determine unknowns $\alpha$,
$\beta$, $\gamma$ of decomposition (\ref{10}) in terms of
derivatives of trajectory. The system (\ref{11}) is overcomplete
because three unknowns are subjected to four equations.

The quantities $\alpha$, $\beta$ are easily expressed from the first
and second equations of the system (\ref{11}),
\begin{equation}\label{gammabeta}
    \gamma=\sqrt{-\alpha}\,,\qquad
    \beta=\frac{B\alpha}{3\alpha-C}\,.
\end{equation}
Substituting this solution into the third and fourth relations
(\ref{11}), we get two polynomial constraints for a remaining
unknown $\alpha$,
\begin{equation}\label{P1}
    P_1(\alpha)=9\alpha^3+9C\alpha^2-(4B^2+4C^2-3E+21)\alpha+BD-EC+7C=0\,.
\end{equation}
\begin{equation}\label{P2}
    P_2(\alpha)=9\alpha^4-6C\alpha^3+(B^2+C^2-9)\alpha^2+6C\alpha-C^2=0\,.
\end{equation}
Relation (\ref{P2}) determines the unknown $\alpha$ in terms of
derivatives of trajectory up to third order. The explicit
representation for $\alpha$ can be found by application of standard
solution to cubic equation (\ref{P1}), for example Cardano formula.
The only negative root is relevant because of condition
(\ref{gammabeta}). We do not provide this solution because the
system (\ref{P1}), (\ref{P2}) admits a simpler representation
without radicals. We give it in the next paragraph. Relation
(\ref{P1}) is another restriction for unknown $\alpha$. Since both
the conditions (\ref{P1}), (\ref{P2}) are consequences of cylinder
equation (\ref{cylinder}), they has to be satisfied simultaneously.
So, $\alpha$ is a common root of polynomials $P_1(\alpha)$ and
$P_2(\alpha)$ . Two different polynomials have a common root if and
only if their resultant with respect to the variable $\alpha$
vanishes,
\begin{equation}\label{Res1}
    \text{Res}_\alpha(P_1(\alpha),P_2(\alpha))=0.
\end{equation}
This is consistency condition for the system (\ref{P1}), (\ref{P2}).
By construction, the resultant is a polynomial in the coefficients
of polynomials (\ref{P1}), (\ref{P2}), being functions of curvature
and torsion of world line, and their derivatives. This resultant is
a differential equation, being satisfied by the cylindrical curves.
It involves derivatives of the path up to fourth order.

Let us now find explicit solution for $\alpha$ and representation
for resultant (\ref{Res1}). Introduce special notation for
combinations of derivatives of curvature and torsion,
\begin{equation}\label{FGH}\begin{array}{c}\displaystyle
    F=-(4B^2+4C^2-3E+21)/9,\qquad G=(BD-EC+7C)/9,\\[5mm]\displaystyle
    H=B^2+C^2-9.
\end{array}\end{equation}
Here, $F$, $G$, $H$ can be considered as alternative combinations
absorbing invariants of trajectory and their derivatives. In terms
of quantities $F$, $G$, $H$ equations (\ref{P1}), (\ref{P2}) take
the most simple form:
\begin{equation}\label{P1-FGH}
    P_1(\alpha)=\alpha^3+C\alpha^2+F\alpha+G=0\,;
\end{equation}
\begin{equation}\label{P2-FGH}
    P_2(\alpha)=9\alpha^4-6C\alpha^3+H\alpha^2+6C\alpha-C^2=0\,.
\end{equation}
The consistency condition in terms of resultant of polynomials
(\ref{P1-FGH}), (\ref{P1-FGH}) reads
\begin{align}
    \notag &\text{Res}_\alpha(P_1(\alpha),P_2(\alpha))=\\[5mm]
    \notag
    &=81F^2G^2H-18FG^2H^2-15C^4F^2H+18C^2F^3H-C^2F^2H^2+486CFG^3-\\[5mm]
    \notag &-324CG^3H+270C^5FG-162C^3F^2G+1458C^2FG^2+18C^3GH+\\[5mm]
    \notag &+189C^2G^2H+4C^4FH-1134C^3FG+15C^2G^2H^2+2C^3GH^2+30C^5GH-\\[5mm]
    \notag &-486CF^3G+G^2H^3-120C^3FGH+90C^2FG^2H-6CFGH^2+108CF^2GH-\\[5mm]
    \notag &-216C^3G+972C^2G^2-36C^4F+504C^5G+540C^4G^2-216C^4F^2+\\[5mm]
    \notag &+36C^6F+540C^3G^3-81C^2F^4-90C^4F^3-1458CG^3+\\[5mm]
    &+C^6H-7C^6+729G^4+15C^8=0.\label{Res2}
\end{align}
The common root of (\ref{P1-FGH}), (\ref{P2-FGH}) can be determined
by the Euclid algorithm. Assuming that this root is simple (i.e. the
curve lies on a single cylinder), we obtain
\begin{equation}\label{alpha-sol}
    \alpha=-\frac{U}{V}
\end{equation}
where
\begin{align}\label{alpha-sol-U}
    \notag & U=225FC^2-6CGH+15C^2FH-216GFG-90C^3G+90C^2F^2-75C^4-\\[5mm]
    &-5C^2H+81G^2-108CG+36C^2+81F^3+FH^2-18F^2H\,;
\end{align}
\begin{align}\label{alpha-sol-V}
    \notag & V=C^3H+GH^2-18FGH+81F^2G-135CG^2-6C^3+15C^2GH+15C^5-\\[5mm]
    &-24C^3F+90FC^2G+99C^2G\,.
\end{align}
Relation (\ref{Res2}) determines the equation of motion for
cylindrical curves. Formulas (\ref{alpha-sol}), (\ref{alpha-sol-U}),
(\ref{alpha-sol-V}) determine a solution to equations (\ref{P1}),
(\ref{P2}) with respect to unknown $\alpha$.

In terms of auxiliary quantity $\alpha$ (\ref{alpha-sol}) the
solution for the momentum $p$ and vector $n$ reads
\begin{equation}\label{p-sol}
    p=\sqrt{-\text{sign}(\sigma)\sigma\varkappa_1\alpha}\bigg(\dot{x}-\frac{B\alpha}{\varkappa_1(3\alpha-C)}
   \ddot{x}
    +\frac{\alpha}{\varkappa_1}[\dot{x},\ddot{x}]\bigg)\,.
\end{equation}
\begin{equation}\label{n-sol}
    n=\frac{\sigma}{\varkappa_1}\bigg(\alpha\ddot{x}+
    \frac{B\alpha}{3\alpha-C}
    [\dot{x},\ddot{x}]\bigg)\,.
\end{equation}
Relations (\ref{pnapva}), (\ref{cylinder}) determine the cylinder
parameters $a$, $v$,
\begin{equation}\label{v-timelike}
    v=\sigma\alpha
    \bigg[\bigg((x,\dot{x})\varkappa_1+(x,\dot{x},\ddot{x})\varkappa_1\alpha-
    (x,\ddot{x})B\alpha\bigg)\dot{x}+
\end{equation}
\begin{equation}\notag
    +\frac{1}{\varkappa_1}\bigg(1-\frac{B\alpha}{C-3\alpha}\bigg((x,\dot{x})\varkappa_1+(x,\dot{x},\ddot{x})\alpha\bigg)-\frac{(x,\ddot{x})B^2\alpha^2}{C-3\alpha}\bigg)\ddot{x}-
\end{equation}
\begin{equation}\notag
    -\frac{1}{\varkappa_1}\bigg((x,\dot{x})\varkappa_1\alpha+(x,\dot{x},\ddot{x})\alpha^2+
    \frac{B}{C-3\alpha}\bigg(1-(x,\ddot{x})\alpha^2\bigg)\bigg)[\dot{x},\ddot{x}]\bigg]\,.
\end{equation}
\begin{equation}\notag
    a=
    \frac{\sigma\alpha}{\varkappa_1}\bigg[\bigg((x,\dot{x})\varkappa_1+(x,\dot{x},\ddot{x})\alpha+
    \frac{(x,\ddot{x})B\alpha}{C-3\alpha}\bigg)^2+
\end{equation}
\begin{equation}\label{a-sol}
    +\frac{(x,\dot{x},\ddot{x})B}{C-3\alpha}-2(x,\ddot{x})\bigg]\,.
\end{equation}
The solution for the total angular momentum $J$ (\ref{J-a}) reads
\begin{equation}\label{J-sol}
    J=\bigg[\frac{(x,p)^2\varkappa_1\alpha-\sigma}{2\sqrt{-\text{sign}(\sigma)\sigma\varkappa_1\alpha}}+\frac{a+\varrho}{2}\sqrt{-\frac{\alpha\varkappa_1}{\sigma}}\bigg]\dot{x}+
\end{equation}
\begin{equation}\notag
+\frac{\alpha}{\varkappa_1}\bigg[\bigg((x,p)+
\frac{B}{C-3\alpha}\frac{(x,p)^2\varkappa_1\alpha+\sigma}{2\sqrt{-\text{sign}(\sigma)\sigma\varkappa_1\alpha}}\bigg)+\frac{B}{(C-3\alpha)}\frac{a+\varrho}{2}\sqrt{-\frac{\alpha\varkappa_1}{\sigma}}\bigg]\ddot{x}+
\end{equation}
\begin{equation}\notag
+\frac{\alpha}{\varkappa_1}\bigg[\bigg(\frac{(x,p)^2\varkappa_1\alpha+\sigma}{2\sqrt{-\text{sign}(\sigma)\sigma\varkappa_1\alpha}}-\frac{B(x,p)}{C-3\alpha}\bigg)+\sqrt{-\frac{\alpha\varkappa_1}{\sigma}}\bigg][\dot{x},\ddot{x}]\,.
\end{equation}
The solution uses auxiliary vector $e$ (\ref{e-p-v}),
\begin{equation}\label{e-sol}
    e=\frac{(x,p)^2\varkappa_1\alpha-\sigma}{2\sqrt{-\text{sign}(\sigma)\sigma^3\varkappa_1\alpha}}\,\dot{x}+\frac{\alpha}{\varkappa_1}\bigg(\frac{(x,p)}{\sigma}+
    \frac{B}{C-3\alpha}\frac{(x,p)^2\varkappa_1\alpha+\sigma}{2\sqrt{-\text{sign}(\sigma)\sigma^3\varkappa_1\alpha}}\bigg)\,\ddot{x}+
    \end{equation}
\begin{equation}\notag
    +\frac{\alpha}{\varkappa_1}\bigg(\frac{(x,p)^2\varkappa_1\alpha+\sigma}{2\sqrt{-\text{sign}(\sigma)\sigma^3\varkappa_1\alpha}}-\frac{B}{C-3\alpha}\frac{(x,p)}{\sigma}\bigg)\,[\dot{x},\ddot{x}]\,.
\end{equation}
Relations (\ref{p-sol}), (\ref{J-sol}) determine the particle
momentum $p$ and particle total angular momentum $J$ (hence, the
cylinder parameters $v$, $a$) in terms of derivatives of classical
trajectory. The description of the particle state is purely
geometrical because no internal variables are involved in
(\ref{p-sol}), (\ref{J-sol}).

Equation (\ref{Res1}) has two obvious gauge symmetries:
reparametrization and translations along the cylinder axis,
\begin{equation}\label{gt-timelike}
    \delta_\xi x=\dot{x}\xi\,,\qquad \delta_\eta x= p \eta.
\end{equation}
(in the last case the vector $p$ is considered as the function of
derivatives of trajectory (\ref{p-sol})). The gauge transformations
are independent for general timelike lines because the vectors
$\dot{x}$, $p$ are non-zero and non-collinear. The vector $p$ is
nonzero for spinning particle with nonzero helicity (see conditions
(\ref{ms-shell})). The velocity vector $\dot{x}\neq0$ is nonzero
because the spinning particle is timelike and casual. The vectors
$\dot{x}$ and $p$ are not collinear because $p$ is lightlike and
$\dot{x}$ is timelike. The tangent space to the cylinder
(\ref{cylinder}) is two-dimensional, so the gauge symmetry is
sufficient to connect each pair of timelike curves that lie on one
and the same world sheet. This result confirms the relationship
between the general timelike cylindrical curves and spinning
particle trajectories.

Now, we can summarize the results of the subsection. We have
associated the timelike spinning particle trajectories as curves on
parabolic cylinders. We have shown that these curves are solutions
to the fourth-order differential equation (\ref{Res1}). Equations
(\ref{p-sol}), (\ref{J-sol}) determine the particle momentum and
total angular momentum in terms of derivatives of trajectory. The
representation for $p$, $J$ does not involve internal variables, so
the particle state description is purely geometrical. Equations
(\ref{gt-timelike}) determine the gauge transformations for spinning
particle trajectories. In section $4$ we show that the geometrical
equation of motion follow from the previously known model. This
ensures that this model describes spinning particle with continuous
helicity at the quantum level.

\subsection{Isotropic paths on the world sheet}\addcontentsline{toc}{section}{Isotropic paths on the world sheet}
Let $x(\tau)$ be an lightlike curve, so the length of the tangent
vector is equal to zero,
\begin{equation}\label{isocond}\begin{array}{c}\displaystyle
    (\dot{x},\dot{x})=0\,.
\end{array}\end{equation}
The natural parameter on the curve (so called pseudo-arc-length) is
determined by the condition
\begin{equation}\label{}\begin{array}{c}\displaystyle
    (\ddot{x},\ddot{x})=1\,.
\end{array}\end{equation}
The natural parameter $\tau$ is well defined on an arbitrary
isotropic curve which is not a straight line. This does not restrict
generality because all the rectilinear paths on the parabolic
cylinder with lightlike axis are cylinder elements, being atypical
curves. The atypical curves are excluded from our consideration.

The Frenet-Serret moving frame, being associated with the isotropic
curve $x(\tau)$, reads
\begin{equation}\label{Frenet-frame-iso}\begin{array}{c}\displaystyle
 e_{0}=\dot{x},\qquad e_{1}=\ddot{x}\,,\qquad
 e_{2}=-\dddot{x}-\frac{1}{2}(\dddot{x},\dddot{x})\dot{x}\,.
\end{array}\end{equation}
The basis $e_a,a=0,1,2$ includes two lightlike vectors $e_0$ and
$e_2$ with normalized scalar product, and normalized timelike vector
$e_1$,
\begin{equation}\label{6-iso}\begin{array}{c}\displaystyle
\phantom{\frac12}(e_{0},e_{2})=(e_{1},e_{1})=1,\quad
(e_{0},e_{0})=(e_{2},e_{2})=(e_{0},e_{1})=(e_{1},e_{2})=0\,.\phantom{\frac12}
\end{array}\end{equation}
The basis (\ref{Frenet-frame-iso}) is well-defined for each
lightlike curve, which is not a straight line.

The Frenet-Serret formulas for the lightlike curve $x(\tau)$ read
\begin{equation}\label{Frenet-formulas-iso}
    \dot{e}_0=e_1\,,\qquad
    \dot{e}_1=\varkappa e_0-e_2\,,\qquad
    \dot{e}_2=-\varkappa e_1\,.
\end{equation}
The lightlike curve is characterized by a single invariant
$\varkappa$, which includes third derivatives of the path,
\begin{equation}\label{7}\begin{array}{c}\displaystyle
\varkappa=-\frac{1}{2}(\dddot{x},\dddot{x})\,.
\end{array}\end{equation}
The quantity $\varkappa$ can be interpreted as some special analog
of curvature, even thought it geometrical meaning is slightly
different. As we will see below, the condition $\varkappa=0$ selects
the lightlike curves on the parabolic cylinder. The representation
of derivatives of trajectory in terms of curvature $\varkappa$ reads
\begin{equation}\label{8}\begin{array}{c}\displaystyle
\dot{x}=e_0\,,\qquad \ddot{x}=e_1\,,\qquad
\dddot{x}=\varkappa e_0-e_2\,,\\[5mm]\displaystyle
\ddddot{x}=\dot\varkappa e_0+2\varkappa e_1\,.
\end{array}\end{equation}
As usual, this representation involves invariants of trajectory up
to fourth order $\varkappa$, $\dot\varkappa$.

We seek for unknown vectors $p$, $n$, being determined by the
conditions  (\ref{3}), (\ref{p-n-cond}) in the following form:
\begin{equation}\label{p-n-iso1}\begin{array}{c}\displaystyle
p=\gamma\sqrt{\text{sign}(\sigma)\sigma}\bigg(\frac{1}{2}\beta^2e_0+\beta
e_1-e_2\bigg)\,,\qquad n=\pm\sigma(\beta e_0+e_1)+\alpha p\,,
\end{array}\end{equation}
where $\alpha,\beta,\gamma$ are new unknowns. The quantity $\alpha$
has the dimension of angular momentum. The quantity $\beta$ has the
dimension of inverse square root of length. The quantity $\gamma$ is
dimensionless. Moreover, we assume $\gamma>0$ in order to meet the
condition $p^0>0$. The sign of $\pm$ determines relative orientation
of vector $p$ with respect to the Frenet-Serret frame
(\ref{Frenet-frame-iso}).

On substituting representation (\ref{p-n-iso1}) into (\ref{3}), we
arrive at the following system of algebraic equations for $\alpha$,
$\beta$ and $\gamma$:
\begin{equation}\label{p-n-iso}\begin{array}{c}\displaystyle
\alpha\gamma=0\,,\qquad
\pm1+\gamma^2+\frac{\alpha\beta\gamma}{\sqrt{\sigma}}=0\,,\qquad
\beta\bigg(\pm1-3\gamma^2-\frac{\alpha\beta\gamma}{2\sqrt{\sigma}}\bigg)+\varkappa\frac{\alpha\gamma}{\sqrt{\sigma}}=0\,,\\[5mm]\displaystyle
2\varkappa\bigg(\pm1+2\gamma^2+\frac{\alpha\beta\gamma}{\sqrt{\sigma}}\bigg)+5\gamma^2\beta^2-\dot{\varkappa}\frac{\alpha\gamma}{\sqrt{\sigma}}=0\,.
\end{array}\end{equation}
The solution to these equations eventually reads
\begin{equation}\label{}
    \alpha=\beta=0\,,\qquad\gamma=1\,.
\end{equation}
For the vectors $p$ and $n$, we find
\begin{equation}\label{p-dddot}
    p=\sqrt{\text{sign}(\sigma)\sigma}\,\dddot{x}\,,\qquad n=-\sigma \ddot{x}\,.
\end{equation}
The cylinder parameters $v$, $a$ read
\begin{equation}\label{pva}
   \phantom{\frac12}v=-\sigma\big(\ddot{x}+(x,\dddot{x})\dddot{x}\big)\,,\qquad
   a=\sigma\big(2(x,\ddot{x})+(x,\dddot{x}){}^2\big)\,.\phantom{\frac12}
\end{equation}
The representation for the total angular momentum $J$ reads
\begin{equation}\label{J-dddot}\begin{array}{c}\displaystyle
    J=-\sqrt{\text{sign}(\sigma)\sigma}\bigg[\dot{x}+(x,\dddot{x})\ddot{x}-\bigg((x,\ddot{x})+\frac{\varrho}{2\sigma}\bigg)\dddot{x}\bigg]\,.
\end{array}\end{equation}
The solution uses explicit representation for the auxiliary vector
$e$ (\ref{e-p-v})
\begin{equation}\label{}\begin{array}{c}\displaystyle
   e=-\frac{1}{\sqrt{\text{sign}(\sigma)\sigma}}\bigg(\dot{x}+(x,\dddot{x})\ddot{x}+\frac{(x,\dddot{x})^2}{2}\dddot{x}\bigg)\,.
\end{array}\end{equation}
The system (\ref{p-n-iso}) has a consistency condition,
\begin{equation}\label{isoeom1}\begin{array}{c}\displaystyle
    \phantom{\frac12}\varkappa=0\,.\phantom{\frac12}
\end{array}\end{equation}
From the differential geometry of the curves, this equation means
that the lightlike path on the parabolic cylinder have zero
curvature. This fact means that the lightlike paths on the parabolic
cylinder with lightlike axis (\ref{cylinder}) are indeed the curves
with zero curvature. The complete set of equations for the
cylindrical paths includes zero curvature condition (\ref{isoeom1})
and the lightlike condition for the particle velocity
(\ref{isocond}),
\begin{equation}\label{EoM-iso}
    (\dddot{x},\dddot{x})=0\,,\qquad (\dot{x},\dot{x})=0\,.
\end{equation}
The first equations in this system has the third order in
derivatives, and the second one has the first order. Equations
(\ref{EoM-iso}) has a single gauge symmetry, being
reparametrization,
\begin{equation}\label{gt-iso}
    \delta_\xi x=\dot{x}\xi\,,
\end{equation}
where $\xi=\xi(\tau)$ is an arbitrary function of proper time.

The identification between the lightlike curves on parabolic
cylinders and the trajectories of spinning particles suggests that
the all the parameters of trajectory are determined by the momentum
and total angular momentum. As we will see, this is not true. The
general solution to equations (\ref{EoM-iso}) reads
\begin{equation}\label{xabcd1}\begin{array}{c}\displaystyle
   x(\tau)=\frac{1}{6}a\tau^3+\frac{1}{2}b\tau^2+c\tau+d\,.
\end{array}\end{equation}
The quantity $\tau$ is a natural parameter on the lightlike curve.
The Cauchy data are constant vectors $a$, $b$, $c$, $d$ that subject
to conditions
\begin{equation}\label{xabcd2}\begin{array}{c}\displaystyle
   \phantom{\frac12}(a,a)=(c,c)=(a,b)=(b,c)=(a,d)=0\phantom{\frac12}\,,\\[5mm]\phantom{\frac12}\phantom{\frac12}\displaystyle -(a,c)=(b,b)=1\,.\phantom{\frac12}
\end{array}\end{equation}
We also assume that $(d,c)=0$. If this is not true, we make a
reparametrization $\tau\mapsto\tau+\gamma$ with appropriate
$\gamma$. The curve (\ref{xabcd1}) lies on the world sheet
(\ref{cylinder}) if
\begin{equation}\label{path-iso}\begin{array}{c}\displaystyle
    a=\frac{1}{\sqrt{\text{sign}(\sigma)\sigma}}p, \qquad b=\frac{1}{\sigma}[p,J],\qquad c=-\sqrt{\text{sign}(\sigma)\sigma}\tau
    e\,,\\[7mm]\displaystyle (d,b)=\frac{1}{\sigma}((J,J)-\varrho).
\end{array}\end{equation}
The particle state determines the parameters of trajectory if these
equations can be solved with respect to $a$, $b$, $c$, $d$. This is
not possible because the solution for $d$ has ambiguity,
\begin{equation}\label{}
    d=\frac{1}{\sigma^2}((J,J)-\varrho)[p,J]+\lambda
    e,\qquad \lambda \in \mathbb{R}.
\end{equation}
The quantity $\lambda$ controls parallel shifts of path along the
symmetry axis of cylinder. It is an additional data, being
independent of $p$ and $J$. Thus the position of general lightlike
cylindrical curve is determined by five parameters. The theory
(\ref{EoM-iso}) cannot be a dynamical system on the continuous
helicity co-orbit (\ref{ms-shell}), which has two physical degrees
of freedom (four physical polarisations).

The results of the subsection demonstrate that the lightlike world
lines are not admissible classical trajectories of relativistic
spinning particle with continuous helicity. In particular, no
geometrical model of continuous helicity spinning particle can be
constructed with lightlike trajectories. The problem of lightlike
curves has no analogue in massive case, where the geometrical models
of relativistic particles with lightlike lines are known for a long
time \cite{NerRam1}, \cite{NerRam2}, \cite{NerMan}. Our no-go result
for continuous helicity particle suggests that the presence of
light-like trajectories is a feature of massive models.

\section{Hamilton's formalism}\addcontentsline{toc}{section}{Hamilton's formalism}

In the articles \cite{GKL1}, the equations of motion of irreducible
spinning particles has been derived from the action functional
involving extra variables, being internal space coordinates. In the
context of current research the model \cite{GKL1} is relevant. The
paper considers the massive particle, but the action functional
admits a smooth continuous helicity limit $m\to0,ms\to\sigma$. In
this section, we demonstrate that the equations of motion for
cylindrical curves follow form the least action principle of the
work \cite{GKL1}. For reasons of simplicity, we consider the case of
spacelike or lightlike spin vector. The accessory parameter
$\varrho$ is determined by the rule
\begin{equation}\label{MM}
    (M,M)=\alpha^2\,,
\end{equation}
which corresponds to the identification $\varrho=\alpha^2$. The case
of lightlike spin vector can be considered in the similar way. We
leave the details to the reader.

The equation of the world sheet can be equivalently rewritten in the
following vector form:
\begin{equation}\label{x-sol}
    x=(x,e)p+(x,p)e-\frac{(x,p){}^2+a}{2\sigma^2}v\,.
\end{equation}
The vector equation has the same valuable information about the
particle position as a scalar relation because it has only one
independent component. The scalar multiplication of left and right
hand sides of equation (\ref{x-sol}) on $p,e$ leads to identity. The
only nontrivial consequence of relation appears after multiplication
of both sides of equation by $v$, and it gives the (\ref{cylinder}).
Relation (\ref{x-sol}) can be considered as the solution to the
world sheet equation in the parametric form. In this setting, the
functions $(x,p)$, $(x,e)$ serve as local coordinates on the
cylinder. Once the spinning particle travels the path on the world
sheet, the quantities $(x,p)$, $(x,e)$ are arbitrary functions of
proper time, being restricted by the causality condition. In what
follows, we assume that the causal trajectories are considered.

Relation (\ref{x-sol}) determines the dynamics of spinning particle.
Differentiating by the proper time, we obtain
\begin{equation}\label{dot-x}
    \dot{x}=(\dot{x},e) p+(\dot{x},p)\bigg(e-\frac{(x,p)}{\sigma^2}v\bigg)\,,
\end{equation}
where $(\dot{x},p)$, $(\dot{x},e)$ are arbitrary functions. The
quantities $(\dot{x},p)$, $(\dot{x},e)$ have sense of velocities of
generalized coordinates $(x,e)$, $(x,p)$ on the world sheet.
Equation (\ref{dot-x}) should be complimented by the conservation
law for momentum and total angular momentum, and the constraints for
the vectors $p$, $v$, $e$,
\begin{equation}\label{dp-v-e}
    \dot{p}=\dot{v}=\dot{e}=0\,;
\end{equation}
\begin{equation}\label{p-v-e-const}
    (p,p)=(e,e)=(e,v)=0,\quad (e,p)-1=0\,,\quad
    (v,v)-\sigma^2=0\,.
\end{equation}
The system (\ref{dot-x}), (\ref{dp-v-e}), (\ref{p-v-e-const})
determine the class of cylindrical curves by obvious reasons.
Equations (\ref{dp-v-e}), (\ref{p-v-e-const}) tell us that the
vectors $p$, $v$, $e$ are integrals of motion subjected to
constraints (\ref{p-v-e-const}). After that the integration of
differential equation (\ref{dot-x}) gives (\ref{x-sol}). The
quantity $a$ appears as the constant of integration. In so doing,
the spinning particle travels along the cylindrical path if and only
if equations are satisfied (\ref{dot-x}), (\ref{dp-v-e}),
(\ref{p-v-e-const}).

The system (\ref{dot-x}), (\ref{dp-v-e}), (\ref{p-v-e-const}) does
not follow from the least action principle of dynamical variables
$x$, $p$, $v$, $e$, $(\dot{x},p)$, $(\dot{x},e)$ because 14
quantities are subjected to 18 evolutionary equations and
constraints. To construct the variational principle, we solve
constraints (\ref{p-v-e-const}) using the lightlike vector $\xi$
with normalized $0$-component,
\begin{equation}\label{}
    \xi=(1,\sin\varphi,\cos\varphi)\,.
\end{equation}
The new dynamical variable $\varphi$ can be considered as angular
variable in the internal space, being a circle. This corresponds to
the configuration space of spinning particle
$\mathbb{R}^{1,2}\times\mathbb{S}^1$. By definition, we put
\begin{equation}\label{v-sol}
    v=\sigma\frac{[\xi,p]}{(\xi,p)}-(\alpha+(x,p))p\,;
\end{equation}
\begin{equation}\label{e-sol}
    e=\frac{\xi}{(\xi,p)}+\frac{(x,p)+\alpha}{\sigma(\xi,p)}[\xi,p]-\frac{((x,p)+\alpha)^2}{2\sigma^2}p\,,
\end{equation}
where $p$ is the momentum, being lightlike vector. Relation
(\ref{v-sol}), (\ref{e-sol}) automatically meets all the constraints
(\ref{p-v-e-const}) involving $v$ and $e$. The only reaming
constraint is the mass shell condition for the particle momentum
\begin{equation}\label{}
    (p,p)=0\,.
\end{equation}

In terms of dynamical variables $x$, $p$, $\xi$ equations
(\ref{dot-x}), (\ref{dp-v-e}), (\ref{p-v-e-const}) take the
following form
\begin{equation}\label{d-xi-x}
    \dot{x}=\bigg((\dot{x},e)+\frac{(x,p)^2-\alpha^2}{2\sigma^2}(\dot{x},p)\bigg)p+(\dot{x},p)\bigg(\frac{\xi}{(\xi,p)}+\frac{\alpha}{\sigma}\frac{[\xi,p]}{(\xi,p)}\bigg)\,;
\end{equation}
\begin{equation}\label{d-xi-v}
    \frac{d}{d\tau}\bigg(\sigma\frac{[\xi,p]}{(\xi,p)}-(\alpha+(x,p))p\bigg)=0\,;
\end{equation}
\begin{equation}\label{d-xi-p}
    \phantom{\frac12\bigg(}\dot{p}=0\,,\qquad (p,p)=0\,.\phantom{\frac12\bigg(}
\end{equation}
(We dot not write out the consequences of the condition $\dot{e}=0$
because the quantity is completely determined by $p$ and $v$.)
Relations (\ref{d-xi-x}), (\ref{d-xi-v}), (\ref{d-xi-p}) have clear
physical sense. Equation (\ref{d-xi-x}) determines the evolution of
the particle position. Equation (\ref{d-xi-p}) tells us that the
vector $p$ conserves, and it is lightlike. Equation (\ref{d-xi-v})
expresses a single independent relation because the vector $v$ has a
single independent component. It has a consequence,
\begin{equation}\label{m-d-xi}
    (\dot{x},p)=\frac{\sigma\dot\varphi}{(\xi,p)}\,.
\end{equation}
On substituting this expression for $(\dot{x},p)$ into
(\ref{d-xi-x}), we obtain the system of two vector equations
(\ref{d-xi-x}), (\ref{d-xi-p}) for the dynamical variables $x$, $p$,
$\varphi$.

Relations (\ref{d-xi-x}), (\ref{d-xi-p}), (\ref{m-d-xi}) follow from
the least action principle for the functional
\begin{equation}\label{S-Ham}
    S=\int\bigg\{ (p,\dot{x})+\frac{\sigma}{(\xi,p)}\dot{\varphi}+
    \alpha\frac{(\partial_\varphi\xi,p)}{(\xi,p)}\dot{\varphi}-\frac{\lambda}{2}
    (p,p)\bigg\}d\tau\,.
\end{equation}
The dynamical variables are the particle position $x$, momentum $p$,
angular variable $\varphi$, and Lagrange multiplier $\lambda$.
Relations (\ref{d-xi-x}), (\ref{d-xi-p}), (\ref{m-d-xi}) appear as
the variational derivatives with respect to the dynamical variables
$x$, $p$, $\lambda$. Taking the Lagrange derivative with respect to
$p$, we obtain equations (\ref{d-xi-x}), (\ref{m-d-xi}),
\begin{equation}\label{}
    \dot{x}=\bigg((\dot{x},e)+\frac{(x,p)^2-\alpha^2}{2\sigma^2}\frac{\sigma\dot{\varphi}}{(\xi,p)}\bigg)p
    +\frac{\sigma\dot{\varphi}}{(\xi,p)}\bigg(\frac{\xi}{(\xi,p)}+\frac{\alpha}{\sigma}\frac{[\xi,p]}{(\xi,p)}\bigg)\,.
\end{equation}
Taking the Lagrange derivative with respect to $x$ and $\lambda$, we
get (\ref{d-xi-p}). The variation of action (\ref{S-Ham}) with
respect to $\varphi$ does not lead to a new independent dynamical
equation because of gauge identity
\begin{equation}\label{}
    (\xi,p)\frac{\delta
    S}{\delta\varphi}+\frac{\sigma\dot{\varphi}}{(\xi,p)}\frac{\delta
    S}{\delta x}=0\,.
\end{equation}
This proves the variational principle for the cylindrical curves.

It remains to verify that the quantization of the classical model
(\ref{S-Ham}) corresponds to the irreducible representation with
helicity $\sigma$ and accessory parameter $\alpha^2$. The fact is
non-trivial because the particles of continuous helicity follow one
and the same paths irrespectively to value of the representation
parameters. The total angular momentum vector reads
\begin{equation}\label{}
    J=[x,p]+\bigg(\frac{\sigma}{(\xi,p)}+\frac{\alpha(\partial_\varphi
    \xi,p)}{(\xi,p)}\bigg)\xi-\alpha\partial_\varphi\xi\,.
\end{equation}
One can see that the vector $J$ meets spin shell condition
\begin{equation}\label{}
    (p,J)=\bigg(p,\bigg(\frac{\sigma}{(\xi,p)}+\frac{\alpha(\partial_\varphi
    \xi,p)}{(\xi,p)}\bigg)\xi-\alpha\partial_\varphi\xi\bigg)\equiv\sigma.
\end{equation}
The spin angular momentum reads
\begin{equation}\label{}
    M=\bigg(\frac{\sigma}{(\xi,p)}+\frac{\alpha(\partial_\varphi
    \xi,p)}{(\xi,p)}\bigg)\xi-\alpha\partial_\varphi\xi\,.
\end{equation}
It is easy to see that condition (\ref{MM}) is true. This result
ensures that the geometrical equations of motion for cylindrical
lines admit equivalent variational formulation with the auxiliary
variables. In its own turn, the variational model can be quantized
in a way that corresponds to the continuous helicity representation
of the Poincare group.

\section{Massless particle}
The massless co-orbit is determined by the relations
(\ref{ms-shell}) with $\sigma=0$. The momentum $p$ and total angular
momentum $J$ of massless particle are subjected to following
mass-shell and spin shell-conditions,
\begin{equation}\label{ml-ms-shell}
\phantom{\frac12}(p,p)=0\,,\qquad (p,J)=0\,.\phantom{\frac12}
\end{equation}
These relations are inconsistent for timelike $J$, so we assume that
the norm of $J$ is nonnegative throughout the section. The spin
vector $M$ is determined by the rule (\ref{Mdef}). Similarly to $J$,
the vector $M$ is lightlike or spacelike. The accessory parameter
$\varrho=\alpha^2$ is determined by the relation
\begin{equation}
\phantom{\frac12}(M,M)=\alpha^2\,.\phantom{\frac12}
\end{equation}

Now, we can discuss the structure of spinning particle world sheet.
Equations (\ref{ml-ms-shell}) have a consequence,
\begin{equation}\label{p-v-sim}
\phantom{\frac12}[J,p]=(J,J)^{\frac12}p\,.\phantom{\frac12}
\end{equation}
It means that the quantity $v$ (\ref{v-a}) and momentum $p$ are
collinear, while the norm of $J$ determines the aspect ratio. With
account of (\ref{p-v-sim}), the equation of the world sheet of
spinning particle eventually reads
\begin{equation}\label{ml-ws}
(J-[x,p])^2-\alpha^2=((x,p)+(J,J)^{\frac12}+\alpha)((x,p)+(J,J)^{\frac12}-\alpha)=0\,.
\end{equation}
The formula determines the pair of parallel hyperplanes with the
normal $p$. The quantity $\alpha$ controls the distance between
hyperplanes. The world sheet is path-connected if $\alpha=0$. In the
latter case, we have a single hyperplane,
\begin{equation}\label{ml-ws-1}
(x,p)+(J,J)^{\frac12}=0\,.
\end{equation}
Here, the vector $p$ serves as the normal, while the norm of $J$
determines the distance between the hyperplane (\ref{ml-ws-1}) and
origin.

Equations (\ref{ml-ws}) and (\ref{ml-ws-1}) tell us that the
positions of massless spinning particle are localized on the
hyperplanes, whose position in Minkowski space is defined by the
values of momentum and total angular momentum.  This fact has been
observed previously in chiral fermion model \cite{Duval1},
\cite{Duval2} earlier. Our result shows that the planar motion has
no alternative for the massless particle irrespectively to any
specifics of the model. In particular, the torsion of the spinning
particle path must be zero in all the instances,
\begin{equation}
    \phantom{\frac12}(\dot{x},\ddot{x}, \dddot{x})=0\,.\phantom{\frac12}
\end{equation}
Unfortunately, this equation contains only partial information about
the model dynamics. The number of independent parameters labelling
the particular hypersurface in the set (\ref{ml-ws}),
(\ref{ml-ws-1}) is less than the co-orbit dimension.  In the case of
two hyperplanes (\ref{ml-ws}), the position of the world sheet is
determined by three parameters: the lightlike vector $p$, and the
norm of total angular momentum $(J,J)^{\frac12}$. In the case of a
single hyperplane (\ref{ml-ws-1}),  only the ratio
$p/(J,J)^{\frac12}$ is relevant. It involves only two initial data
in independent way. The dimension of co-orbit (\ref{ml-ms-shell})
equals four in all the instances. The extra dynamical degrees of
freedom have no geometrical description in terms of world sheet
formalism,  and they require introduction of internal variables.
This result means that world sheet concept cannot be used for
construction of geometric model of massless spinning particle. On
the other hand, the spinning particles must be planar curves in
every irreducible spinning particle theory. As it has been mentioned
above, this condition is satisfied for previously known models.

\section{Conclusion}

In the current article, the recently proposed idea of characterising
the classical spinning particle dynamics by the world sheet, rather
than the world line \cite{LK}, has been applied to the problem of
description of dynamics of irreducible spinning particle with
continuous helicity. It has been shown that the admissible classical
positions of the particle lie on parabolic hypercylinder in
Minkowski space irrespectively to any specifics of the model. The
position of the hypercylinder is determined by the values of
momentum and total angular momentum. The focal distance is
determined by helicity. The classical trajectories of the spinning
particle are given by causal cylindrical lines. Assuming that all
the trajectories belonging to the same cylinder are connected by
gauge transformation, we have derived an ordinary differential
equation describing the general cylindrical lines with time-like
tangent vector. These equations of motion are purely geometrical,
and they involve invariants of the classical path including the
derivatives up to the fourth order. The momentum and total angular
momentum are expressed as the functions of trajectory. To our best
knowledge, geometrical equations of motion, not involving any extra
variables, have been previously unknown for the continuous helicity
particle.

We have paid the particular attention to the class of lightlike
cylindrical lines. It has been shown that the cylindrical lightlike
trajectory either a straight line (representing the cylinder
element) or the curve with zero (lightlike) curvature. Unlike the
massive case, no lightlike curves can serve as physically acceptable
trajectories of spinning particles. The lines lie on infinite number
of parabolic cylinders with one and the same direction of axis.
These trajectories do not determine the state of the particle in
unambiguous way. As for zero curvature paths, their position in
spacetime involves besides the momentum and total angular momentum
an extra initial data. Having extra degree of freedom, the theory of
lightlike cylindrical curves cannot be considered as the a dynamical
system on the continuous helicity co-orbit. As the spinning
particles are dynamical systems on the co-orbit, this theory can't
describe the motion of spinning particle.

We have proven that the geometric equations of motion for
cylindrical curves can be derived from the least action principle.
This is important from several viewpoints. First, the concept of
irreducible spinning particle suggests that the model can be
quantized, while its quantization corresponds to the irreducible
representation of the Poincare group. The variational principle
provides the way to constructing the quantum theory. Second, the
variational principle shows that our results are consistent with the
previous studies. We explicitly demonstrate that the differential
equations for cylindrical curves follow from the action functional
of the work \cite{GKL1}, and vice versa.  The quantization of this
model do correspond to the continuous helicity representation. In
all the cases, the variational principle involves extra variables
having sense of coordinates in internal space.

The studies of the spinning particle world sheet concept can be
continued in several directions. One of the interesting issues is
the geometry trajectories of spinning particle in the external
electromagnetic and gravitational field. The article \cite{KR1}
tells us that the world sheet of massive particle in electromagnetic
field is a cylindrical hypersurface, whose radius is fixed by the
representation. The world sheets of continuous helicity may have
much more interesting geometry because the sections of parabolic
cylinder are not compact. The couplings between the particle
traveling a cylindrical path and external field are expected to be
non-local, while the non-locality is controlled by the helicity.
Expanding these equations in helicity, we will obtain approximate
equations describing cylindrical trajectories of continuous helicity
spinning particles. The leading orders of these equations will serve
as the analogs of Frenkel \cite{Fren} and Mathisson-Papapetrou
\cite{Mathisson}, \cite{Papapetrou} models .

\section*{Acknowledgments}
The authors thank A.A. Sharapov for valuable discussions of this
work. The work was supported by RFBR (project number 20-32-70023)
and Foundation for the Advancement of Theoretical Physics and
Mathematics ``BASIS".

\renewcommand\refname{Bibliography}

\end{document}